\documentclass[aps,prl,onecolumn, aarticle, superscriptaddress, ]{revtex4-1}
\usepackage{graphicx}
\usepackage{float}
\usepackage{multirow}
\usepackage{array}
\usepackage{caption}
\usepackage{subcaption}

\begin{document}


\title{Variation of Instructor-Student Interactions in an Introductory Interactive Physics Course}


\author{Emily A. West}
\noaffiliation{}
\author{Cassandra A. Paul}
\affiliation{Department of Physics \& Astronomy, San Jos\'e State University, One Washington Square, San Jos\'e, CA 95192}
\author{David Webb}
\author{Wendell H. Potter}
\affiliation{Department of Physics, University of California at Davis, One Shields Ave, Davis, CA 95616, USA}
\date{\today}
\begin{abstract}
The physics instruction at UC Davis for life science majors takes place in a long-standing reformed large-enrollment physics course in which the discussion/lab instructors (primarily graduate student teaching assistants) implement the interactive-engagement (IE) elements of the course.  Because so many different instructors participate in disseminating the IE course elements, we find it essential to the instructorsÕ professional development to observe and document the student-instructor interactions within the classroom.  Out of this effort, we have developed a computerized real-time instructor observation tool (RIOT) to take data of student-instructor interactions.  We use the RIOT to observe 29 different instructors for five hours each over the course of one quarter, for a total of about 150 hours of class time, finding that the range of instructor behaviors is more extreme than previously assumed.  In this paper, we introduce the RIOT and describe how the variation present across 29 different instructors can provide students in the same course with significantly different course experiences.
\end{abstract}
\maketitle
\section {Introduction and Background}


Physics education researchers have shown in numerous studies that interactive-engagement (IE) courses promote student learning of physics concepts better than their traditional counterparts \cite{Hake1998IE, Prince, Koenig2007Effectiveness}. These courses are defined as those in which students engage in activities where they make sense of material while engaging in discussion with peers and/or instructors.  Because IE courses are for the most part successful at improving student learning outcomes, many physics departments nationwide are adopting IE courses
 \cite{Tutorials, Potter96, Webb}. 
However, there is a large variety in the success of interactive-engagement courses \cite{Hake1998IE}. For some IE courses the student conceptual understanding gains, as measured by the Force Concept Inventory (FCI)\cite{FCI},
are indistinguishable from those in traditional courses. 
Additionally, while traditional courses typically produce normalized FCI gains tightly clustered around an average of .23, an interactive-engagement course can produce gains anywhere form .23 to .70! \cite{Hake1998IE}

Some of the variation in IE course success is due to the broad curricular definition of the term Òinteractive-engagement,Ó but a portion of this variation is also likely due to the pedagogical implementation at the instructor (faculty or graduate student) level  \cite{Turpen2009All, Paul2010}. That is to say that some of the variation is due to differences in course structure (time devoted to different types of content, balance between lecture and lab and discussion, etc.) whereas other aspects of the variation are due to the manner in which a particular instructor interacts with the class. 

It is typical at the university level for many instructors to teach different sections of the same course. This is particularly true for graduate student teaching assistants (TAs).  As more IE courses are being implemented\cite{Koenig2007Effectiveness, ScaleUp, Chance}, graduate students are becoming increasingly responsible for using interactive pedagogy because often the IE elements are present in the labs and discussion courses.

As of yet, there has been little research conducted with regard to graduate teaching assistant pedagogy. That which does exist suggests that there is a variation in types of interactions the instructors engage in.  A former member of the UC Davis physics education research group completed dissertation research on TA actions in the classroom, defining several different typical instructor modes ranging from ÒobserverÓ to ÒobtrusiveÓ\cite{Austin}. Additionally, Goertzen has found that not all TAs have the same amount of Ôbuy-inÕ to course philosophy, which can affect how they implement the course material \cite{Goertzen2009Accounting}. She also has found evidence that TAs use different indicators to determine if their students are making sense of the material \cite{Goertzen2008Indicators}. Finally, emerging research shows that these differences, specifically instructorsÕ responses to student questions, can stifle or promote student reasoning \cite{KarelinaEtkina}.  These findings suggest that students can have vastly different experiences in different sections of the same course taught by different instructors. 

There are ways to measure the variation of instructor implementation, for example, the ÒReformed Teaching Observation Protocol (RTOP)Ó was developed to measure the extent to which interactive and student centered techniques are utilized in a given classroom \cite{RTOP}. The RTOP is a twenty-five item instrument that allows an observer to rate the extent to which particular IE elements are present in a lesson. The RTOP is an excellent tool for evaluating reformed classrooms, however, the RTOP assumes that the instructor is also the curriculum/lesson designer, which is not always the case.  In higher education, often graduate student teaching assistants are teaching the IE component (as in the popular Washington Tutorials for example). Furthermore, the overall RTOP score does not give us an illustrative view of the classroom, and while field notes may accomplish this to some extent, it is limited to what the observer decides to write down.
 

We currently lack both a feel for the pedagogical variation that exists in IE courses, and a tool that illustrates the difference. Our research fills an important gap in the literature by describing the teaching practices of 29 instructors in a long-standing reformed course supplemented by an extensive TA training and professional development experience for the graduate student instructors. In this paper we examine the pedagogical landscapes that exist in an IE course where all instructors guide the students through identical activities, with identical sets of instructor notes.  We use the ÒReal-time Instructor Observing Tool (RIOT),Ó a computer program that allows us to code during actual classroom observations, to time and quantify the instructor-student interactions in the classrooms of the twenty-five TAs and four faculty members teaching the long-standing introductory IE course at UC Davis.

Defining the pedagogical variation among instructors in a course where the curriculum and student population are constant is the first step in determining if this difference has an impact on student learning outcomes.

\section {Sample and Environment}
Observations take place in a large-enrollment three-quarter introductory physics course for non-majors at UC Davis with more than 1500 students enrolled each academic year.  The course  \cite{Webb, Potter96, Potter00, Potter01, Deleone00} designed and institutionalized in the mid 1990s in coordination with educational research findings, and has previously been presented and published simply by referencing the course name, ``Physics 7."  The course design has recently been renamed as Collaborative Learning through Active Sense-making in Physics (CLASP) to more meaningfully capture the course approach and goals. The class is nominally calculus-based, though calculus is used sparingly throughout the course.  All students at UC Davis taking this level of physics enroll in a CLASP course; there is no traditional alternative so these students are not a selective group.  In a CLASP course, students are expected to use model-based reasoning \cite{Hestenes87, Halloun87, sep-models-science, models02} to Ômake senseÕ of physical phenomena.  This sense-making is done in an interactive ``Discussion/Lab" (DL) which is an environment where 25-30 students work in small groups of five at tables and blackboards.  While the inclusion of model-based reasoning activities is one of the elements of the course that is believed to make it effective \cite{Webb, Potter96, Deleone00}, this paper will concentrate on the interactive-engagement aspect of the course. Students in the course spend five hours per week in bi-weekly discussion/labs, while only attending lecture 80 minutes a week (25 minutes of which are typically reserved for weekly quizzes).  

The CLASP discussion/labs are where the students are expected to learn the bulk of the course content.  Learning takes place through a series of activity cycles in which student group work is spliced with instructor led ÒWhole Class Discussions.Ó  During the small group time, students are expected to work together to respond to activity prompts on their group's blackboard.  The instructor monitors progress of the whole class simultaneously by scanning the blackboards and can then identify students having trouble.  When assisting, instructors are not taught to provide immediate answers, but to help guide students back into a productive thinking space until the students can construct a logical argument explaining the physics of the current activity.  The instructor is encouraged to conduct a whole class discussion after some groups have finished and all groups have made significant progress toward a complete argument. During whole class discussions, the instructor and students work together to discuss the major points of the activity and larger implications of the phenomena.  At this time instructors may call on individuals or groups to present their work, ask follow up questions to test for understanding, guide the class to a synthesis of the larger implications of the model, or any combination of these approaches.  Ideally the instructor facilitates the discussion, but the students voice the majority of the ideas.  In cases of class-wide confusion, students should feel comfortable asking questions and participating actively in the discussion.  In both small group time and whole class time, the instructors play the role of guiding the students towards correct use of the models. The above description captures the philosophy of those who created the CLASP curriculum; this study shows the extent to which the philosophy is implemented in practice.

The instructors who teach CLASP DL sections are mostly graduate students, though each quarter a few temporary and/or permanent faculty also lead DL sections. While no instructor who strongly dislikes the CLASP style is forced to teach CLASP courses, approximately half (30 out of 63) of the graduate student teaching positions available are for CLASP courses.  All new graduate student TAs (which includes the vast majority of all first year graduate students) teach one of the three CLASP courses on arrival.  Of the 30 instructors in our study, 26 are graduate students. Thirteen of the graduate TAs are in their first year and 13 are in their second year or beyond. Two of the 26 graduate students are associated with the PER group.  All graduate students serving as DL instructors experience a two-part department-run professional development course.  Before teaching, instructors participate in an intense two-day training course.  Then throughout their first quarter, professional development continues one hour per week, including discussions of pedagogy and several opportunities for new TAs to observe one another and experienced DL instructors.  The professional development is aimed at communicating the course philosophy to the graduate DL instructors.  The details of these programs are not important to the current discussion, though understanding that all instructors have experienced considerable training and exposure to the course philosophy is important.  
Four of the DL instructors who taught CLASP discussion/labs in winter 2008 hold PhDs and are permanent or temporary lecturers or faculty.  These instructors are not required to attend any training or professional development associated with CLASP but two of these individuals are actively involved in the physics education research group at UC Davis.

\section{Methods}
Our study focused on observing the interactive elements of CLASP, so we focused only on the instructors who taught the discussion/labs and did not include observations of the 80 minute/week lectures. Each of the 30 instructors in our study was observed twice during the quarter.  Every observation lasted the entirety of the DL (approximately 140 minutes).  The researcher did not interact with the instructor or the students during the observation, and no feedback was given to the instructors about their teaching practice until the quarter was over and all forms of data collection were completed. 

The primary data were gathered using a computerized tool called ÒRIOTÓ (Real-time Instructor Observing Tool).  At the most basic level, the observation protocol differentiates between four major types of interactions; talking at students, shared instructor-student dialog, observing students, or not interacting.  These original types of interactions were chosen years ago to capture TA behaviors of interest to the CLASP course designers.  Some form of the RIOT has been used during professional development for years at UC Davis.  The original observation protocol required a once per minute observation of the instructorÕs action at that instant, ignoring the other 59 seconds.  Feeling that this eliminated much useful data, we retooled the protocol to use real-time observations through a series of scripts within File Maker Pro database software.  During data collection, observers operate on a self-imposed 10 second delay.  That is, when an instructor switches from one type of interaction to another, the observer waits 10 seconds before recording the change to make certain that the type of interaction is understood.  Any observation taking less than 10 seconds is ignored on the principle that it will not have a lasting impact on the resulting classroom culture.  
Before beginning data collection, we also subdivided the major interaction types into smaller categories to more richly describe the types of interactions within the classroom.  These categories were chosen during preliminary observations and emerged from the data we collected.  A summary of all categories of interactions is shown in Table~\ref{tab:interactions}.  The instructor can perform each category of interaction with an individual student, one of five or six different small groups, or with the whole class.  Each different combination (category and with whom the interactions are with) appears in a two-dimensional color-coded grid of buttons on the laptop screen.  By design, the instructor can only engage in one interaction category at a time.  A screen shot of the program interface is included in the appendix (Fig.~\ref{fig:screenshot}).

\begingroup 
\squeezetable 
\begin{table}
 \caption{A list of all possible Instructor-Student interactions captured by the RIOT.}
 \label{tab:interactions}
\begin{tabular}{|p{4cm}|p{4.1cm}|p{8cm}|} \hline

\multirow{1}{*}{{\bf Type of Interaction}} & {\bf Category of Interaction} & {\bf Description} \\ \hline
\multirow{2}{*}{Talking at Students}
& \multicolumn{1}{l|}{Explaining} & Instructor is explaining physics concepts, answers, or processes to student(s)\cr \cline{2-3} & Clarifying Instructions & Instructor is clarifying the instructions, or reading from the activity sheet, or covering logistical issues, or transitioning between SG and WCD\cr \hline

\multirow{4}{4cm}{Dialoguing with Students}
& \multicolumn{1}{l|}{Listening to Question} & Instructor is listening to a student's question\cr \cline{2-3} 
& Engaging in Closed Dialogue & Instructor is controlling conversation, but not explaining. Most commonly, a series of short questions meant to lead the student to the correct answer. (ex: Was energy lost or gained by the system? So is this an open or closed system?)  Student contribution is one to several words at a time. \cr \cline{2-3}
& Engaging in Open Dialogue  \footnotemark[1]& Original definition: Instructor is participating in a student led conversation.  Student may be leading instructor through thinking that had occurred previously or might be actively Òmaking senseÓ with the instructor and classmates.  Student contribution is complete sentences.
Refined definition: Students are contributing complete sentences, though not actively Òmaking sense.Ó (sense-making moved to ÒideasÓ, below)
\cr \cline{2-3}
& Engaging in Open Dialogue with Ideas being shared  \footnotemark[1] & Instructor is participating in student led conversation.  Student specifically is having a productive conversation with the instructor where concepts are being challenged and worked on.  Student contribution is complete sentences, specifically the active development of physics ideas. \cr \hline

\multirow{4}{*}{Observing} 
& \multicolumn{1}{l|}{Passive Observing} & Instructor is scanning room and assessing classroom progress from afar or browsing blackboard work of groups for less than 10 seconds at a time.\cr \cline{2-3} 
& Active Observing & Instructor is actively listening to individual students or groups. \cr \cline{2-3}
& Student Presentation & Instructor is listening to students presenting their work to the class. \cr \cline{2-3}
& Students Talking Serially & Instructor is listening to students talking serially in a whole class discussion.  Students are asking each other questions and building on each otherÕs ideas.  Note: Although we think this is an important event, it happened rarely, and therefore was not considered in the analysis. \cr \hline

\multirow{5}{*}{Not Interacting}
& \multicolumn{1}{l|}{Administrative/Grading} & Instructor is grading student homework, or discussing quizzes, lateness or other policy.\cr \cline{2-3}
& Working on Apparatus & Instructor is helping students with experimental apparatus or computers.\cr \cline{2-3}
& Chatting & Instructor is chatting socially with students.\cr \cline{2-3}
& Class Prep & Instructor is reading notes, or writing something on the board while the students are in their small groups.\cr \cline{2-3}
& Out of room & Instructor has left the room. \cr \hline

\end{tabular}

 \footnotetext[1]{After nearly half the observations were completed (24 of the 30 were observed once), two additional categories were needed.  Because this study involves coding in real time, we could not re-categorize the data we took previously.  One category we added was "Chatting" socially with students.  Before adding this category, any chatting was previously categorized under "Administrative/Grading."   The ÔideasÕ category was added to split the Ôopen dialogueÕ category by distinguishing between students leading the instructor through calculations (unit conversions for example), and students thoughtfully stringing ideas together with the instructor and classmates.  The 'ideas' category is essentially a type of 'Open Dialogue.'  To clarify, for 24 of the observations, ÔideasÕ is contained in Ôopen,Õ and for 35 observations, ÔideasÕ is its own category.}

\end{table}
\endgroup

All observations conducted during the study were coded using the RIOT by one of two researchers.  Prior to the start of the study, the two researchers observed the same classroom on three occasions in order to establish inter-rater reliability.  Between each observation, the two collaborated extensively to establish a reliable definition of what each category included and excluded.  In order to ensure consistent coding, the observers also coded the same classroom for approximately 45 minutes every week.  They continuously referred to the category definitions and updated them as necessary. For the length of time corresponding to an entire DL period, 140 minutes, inter-rater reliability varied between 85 and 95 percent.  

The main CLASP course series is three quarters long with parts A, B, and C.  During the quarter of observation, five independent CLASP courses were taught: two A and two B courses, but only one C course.  In each of the ~300-student courses there are 11 discussion/lab sections, typically with five (5) TAs teaching 10 of the 11 DL sections and a faculty member/lecturer teaching one section. In each of the five sections (A-1, A-2, B-1, B-2, or C) every instructor was observed during the same DL meeting, enabling direct comparison to all other observations that covered the exact same activities, with two exceptions. See Table 2.  In addition, it was possible to observe the same DL across the entire sample of all 12 CLASP A instructors (6 from each session).  This was also true for CLASP B observations, though not for CLASP C since only one session of C occurred during the quarter of the study.   Observing all instances of the same DL meeting (across both sessions of A and across both of B) was exhausting to accomplish because they occurred during the same week, so it was only done for one of the two ÔsetsÕ of observations; DL 11 for A and DL 13 for B (see TABLE~\ref{Course}).
 \begin{table}
 \caption{Course and Observation Breakdown.\label{Course}}
 \begin{ruledtabular}
 \begin{tabular}{cccc}
 Course Segment & Lecture Session & Number of DL Instructors & DL Meeting Observed\\
 CLASP A & 1 & 6 & 6, 11\\
 CLASP A & 2 & 6 & 11, 17, 16 \footnotemark[1]\\
 CLASP B & 3 & 5 \footnotemark[2]& 4, 13 \\
 CLASP B & 4 & 6 & 3, 13 \\
 CLASP C & 5 & 6 & 7, 13 \\
 \end{tabular}
 \end{ruledtabular}
 \footnotetext[1] {One instructor was observed during DL 16 instead of DL 11 due to their having a substitute on the planned observation day.}
 \footnotetext[2] {The 6th instructor was dropped from the study because the person teaching the section changed mid-quarter.}
 
 \end{table}


 \begin{figure*}[t!]
   \subcaptionbox{\label{fig:T3a}}
{\includegraphics[height=.22\textheight]{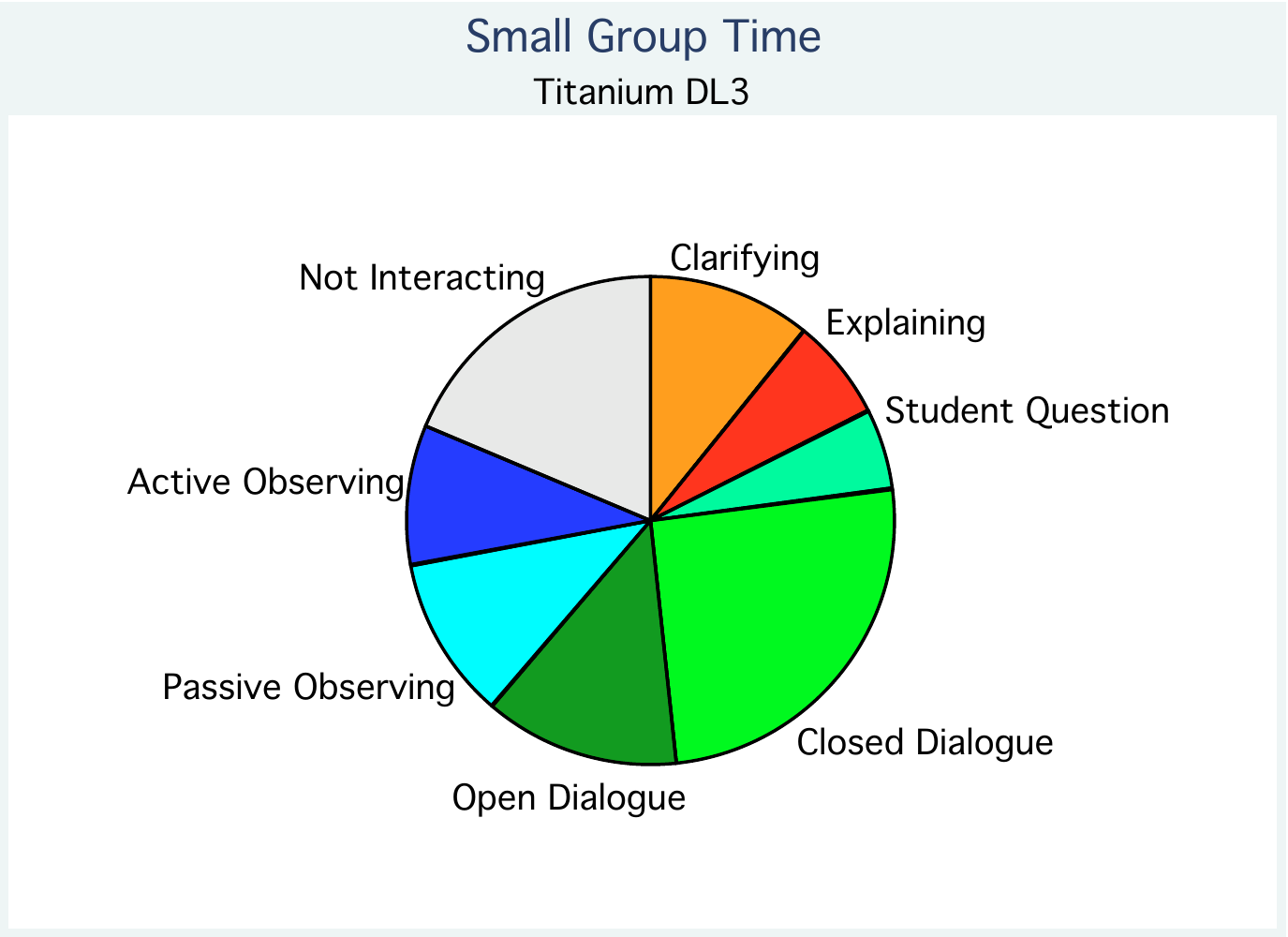}}
 \subcaptionbox{\label{fig:T3b}}
 { \includegraphics[height=.22\textheight]{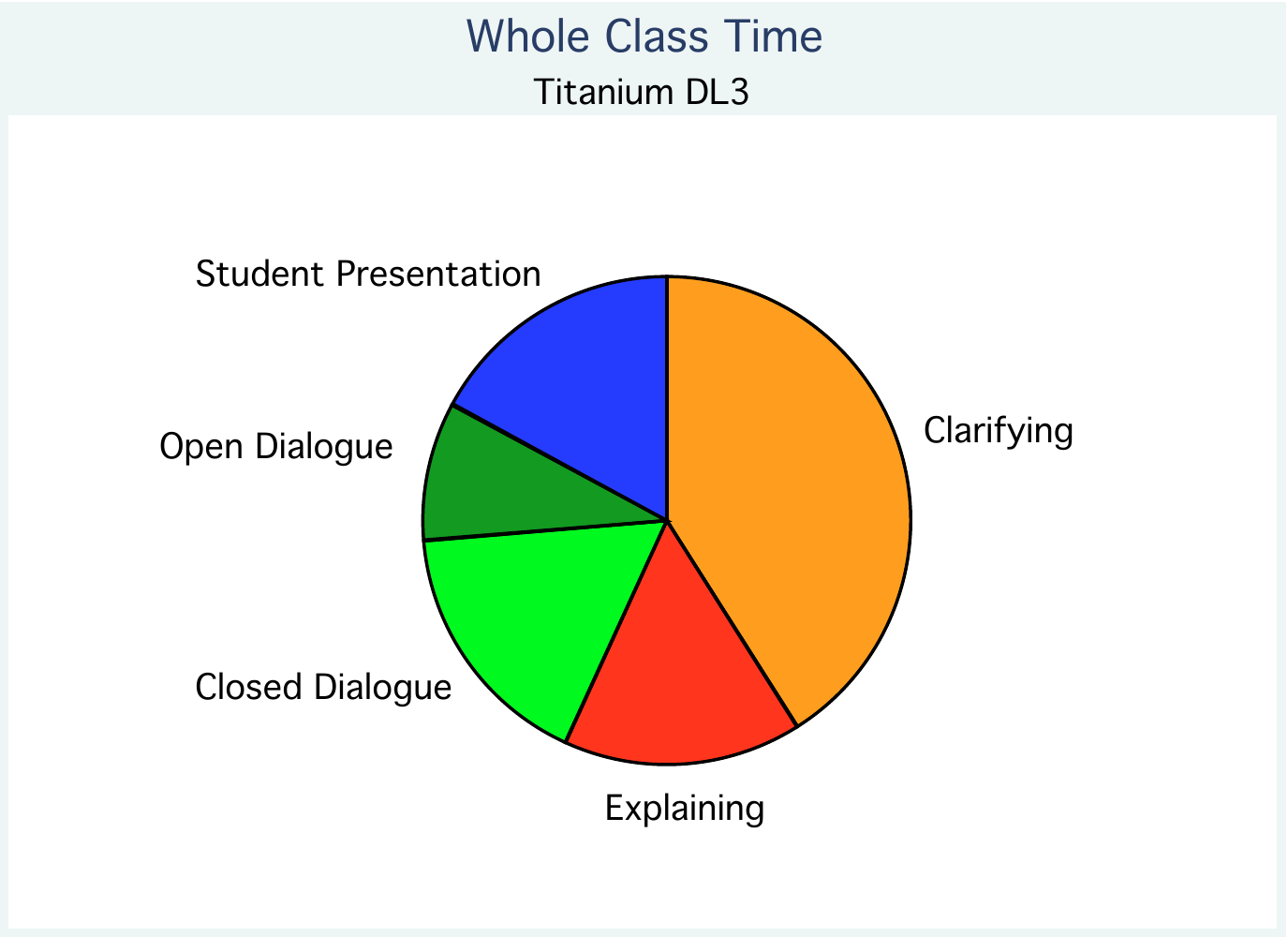}}
\subcaptionbox{\label{fig:T3c}}
   { \includegraphics[height=.18\textheight]{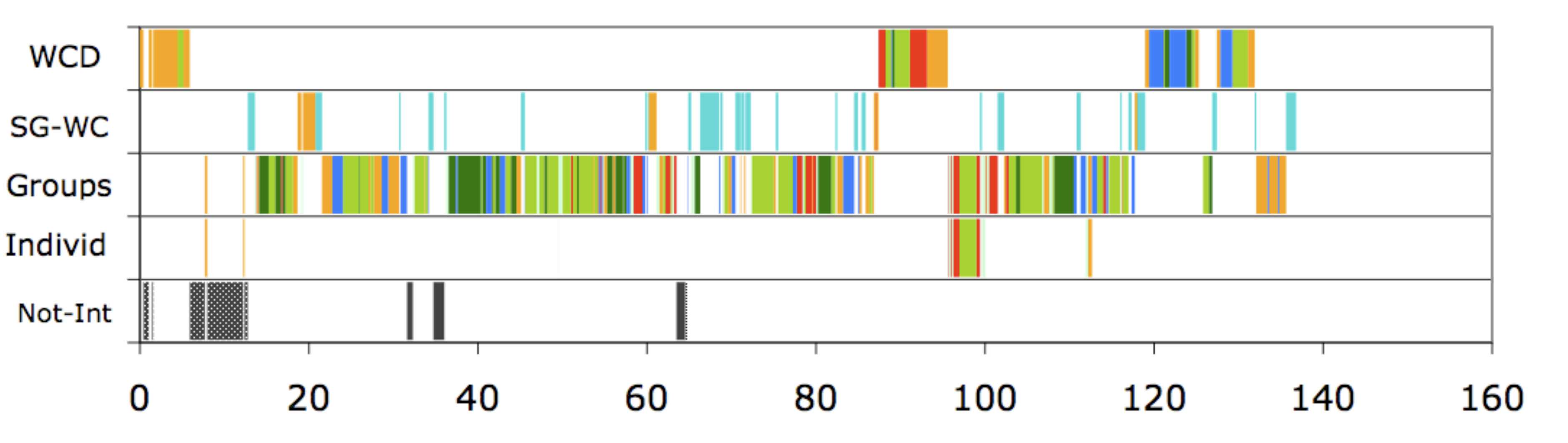}}

 \caption{Instructor Titanium (DL 3, CLASP B): Small Group time breakdown (a), and Whole Class breakdown (b), and the time series data (c).  The SG-WC line refers to time when the instructor is addressing the whole class during Small Group time.  By definition, Passive Observing falls here, but there are other times when the instructor needs to address the whole class briefly without calling for a Whole Class discussion.  Compare with Figure~\ref{fig:Li3}, both instructors are teaching the same set of activities.  Notice that they spend their time interacting in similar ways during the Small Group time, but have a very different set of interactions during Whole Class time.}
 \label{fig:T3}
 \end{figure*}

\section{Results}
\begin{figure*}[t!]
   \subcaptionbox{\label{fig:Li3a}}
 {\includegraphics[height=.22\textheight]{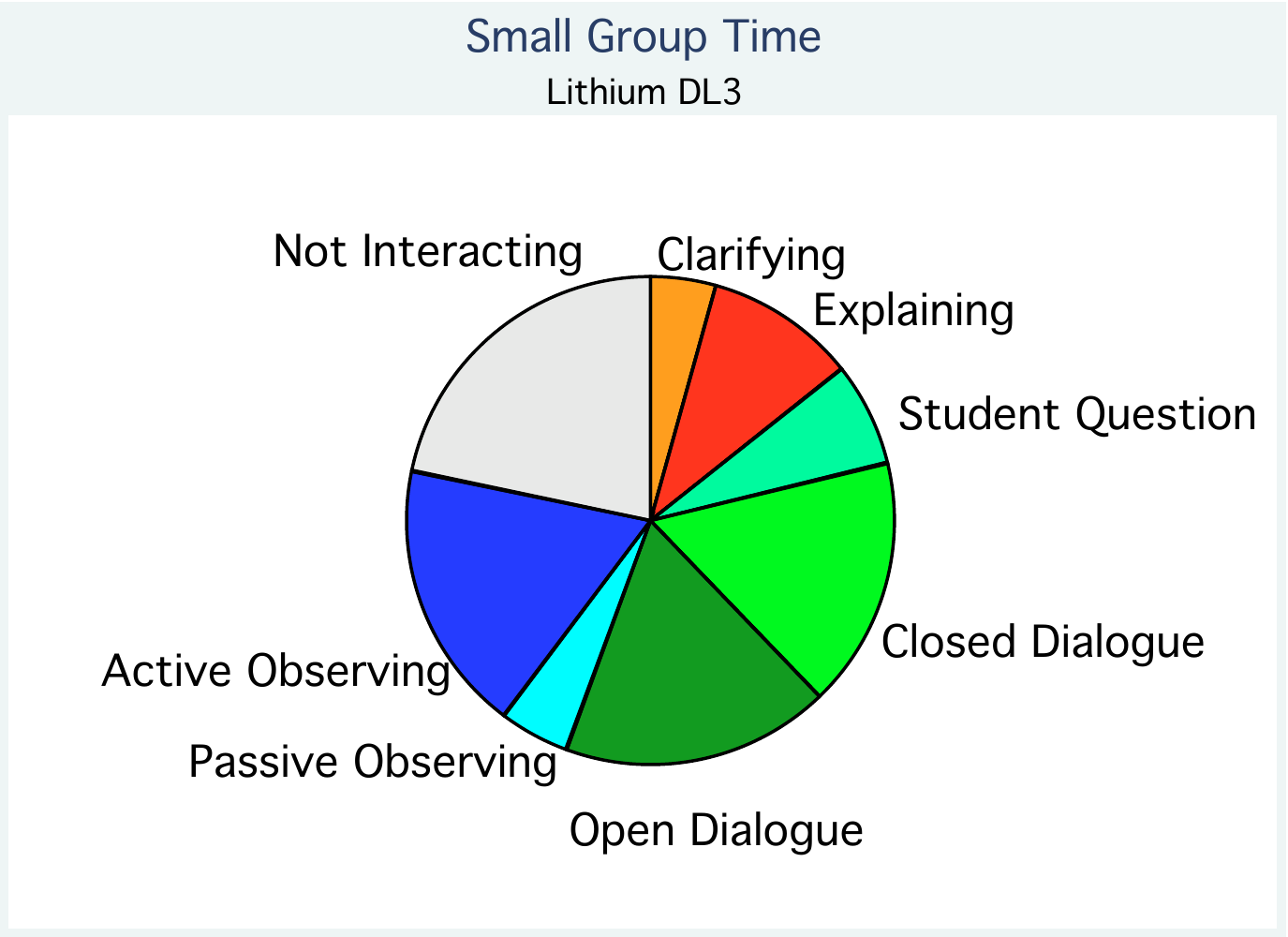}}
   \subcaptionbox{\label{fig:Li3b}}
  {\includegraphics[height=.22\textheight]{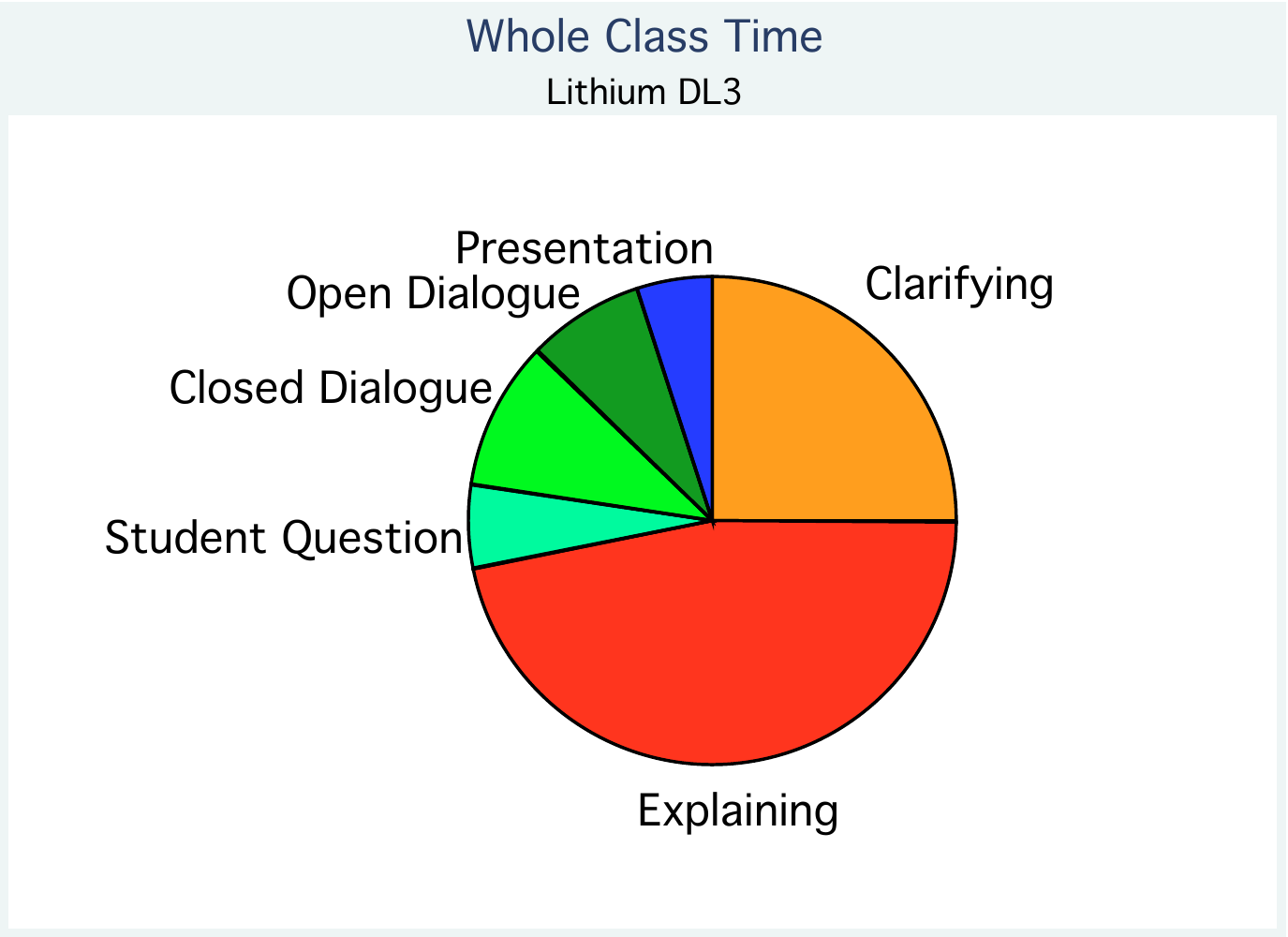}}
     \subcaptionbox{\label{fig:Li3c}}
   { \includegraphics[height=.18\textheight]{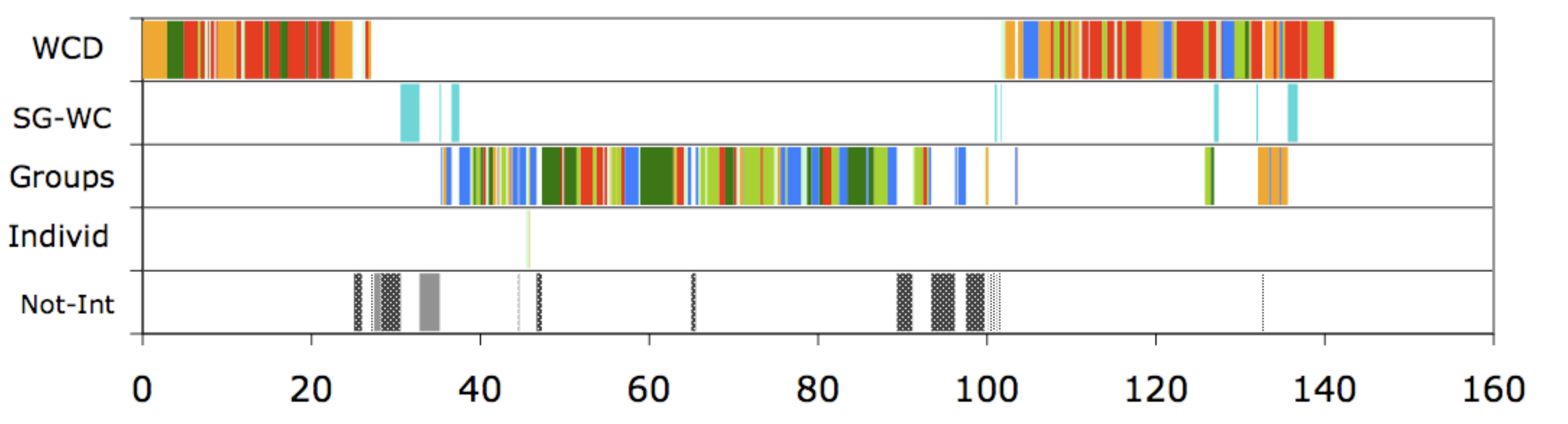}}
 \caption{Instructor Lithium DL 3: Small Group time breakdown (a), Whole Class breakdown (b), and the time series data (c).  The SG-WC line refers to time when the instructor is addressing the whole class during Small Group time.  By definition, Passive Observing falls here, but there are other times when the instructor needs to address the whole class briefly without calling for a Whole Class discussion. Compare with Figure~\ref{fig:T3}, both instructors are teaching the same set of activities.  Notice that they spend their time interacting in similar ways during the Small Group time, but have a very different set of interactions during Whole Class time.}
 \label{fig:Li3}
 \end{figure*}

In order to maintain anonymity, each instructor in our study was given a code name corresponding to one of the first 30 elements.   Our results reveal a great deal of variation in instructor-student interactions.   This variation occurs in the relative amount of small group (SG) versus whole class discussion (WCD) time, the type of interactions that occur during SG time, and the type of interactions that occur during WCD time. Additionally, we find a great deal of independence between the distribution of interactions in SG and WCD time.  That is, instructors with similar distributions of interactions in SG time did not necessarily have similar distributions in WCD time, and vice versa  (compare figures~\ref{fig:T3} \&~\ref{fig:Li3} and appendix ~\ref{fig:7Aobs1} \& ~\ref{fig:7Bobs2}).  As a result, we have analyzed the two separately.  A larger sample of our data is available in the appendix.

\subsection{WCD vs. SG allocation}
\begin{figure*}[t!]
   \subcaptionbox{\label{fig:breakdowna}}
 {\includegraphics[height=.25\textheight]{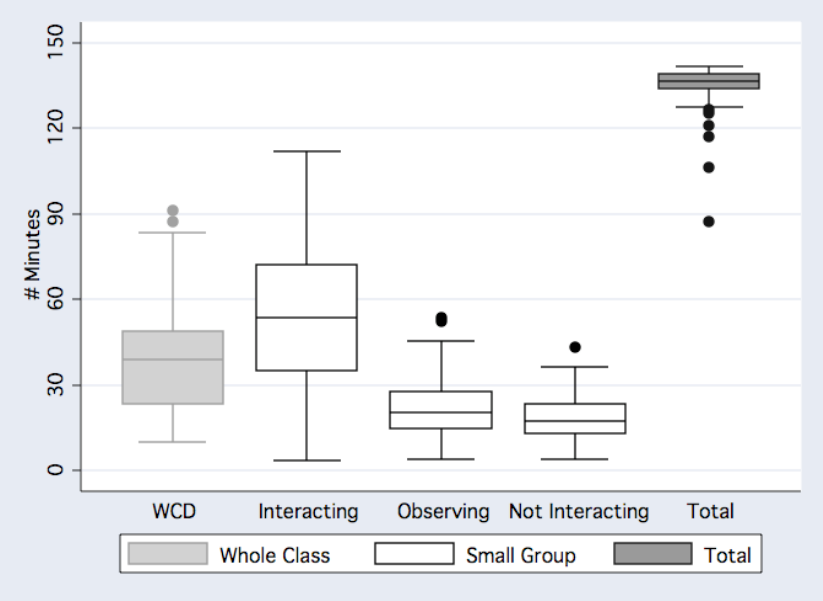}}
   \subcaptionbox{\label{fig:breakdownb}}
  {\includegraphics[height=.25\textheight]{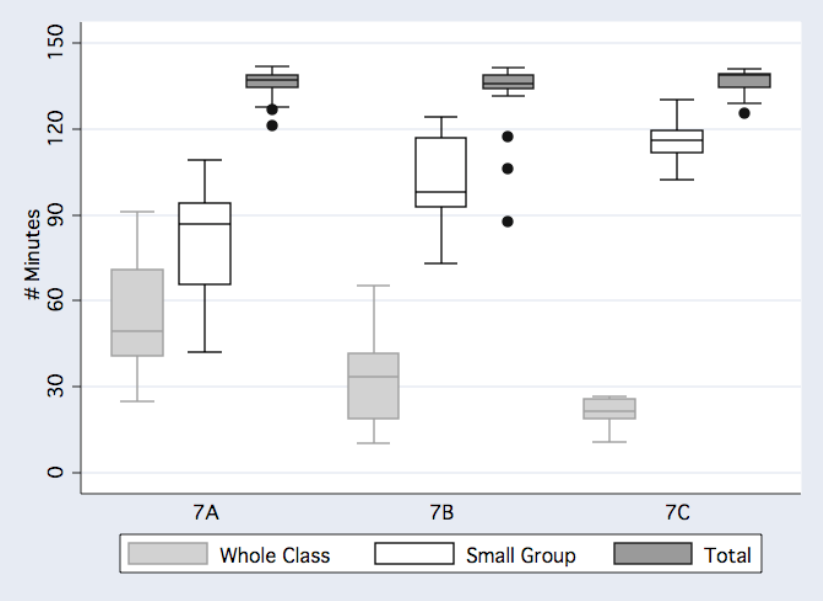}}
 \caption{Breakdown of time between whole class discussion and small group time.  Times are in minutes.  A complete allotted DL period is 140 minutes.  The box contains values from the 25th to 75th percentile, with the median marked.  The outliers are shown as dots, not connected to the box. (a) Across the whole sample, small group time is further divided into time spent interacting with students, observing students, and time spent not interacting.  (b) The overall breakdown of time between WCD and SG time is broken down by course.}
 \label{fig:WCSGtime}
 \end{figure*}

First, we present the overall time allocation in discussion/lab for all 59 observations (see Figure~\ref{fig:WCSGtime}).  The average DL section is released five minutes early (or started 5 minutes late), taking 135 of the total possible 140 minutes, and there is little variation in the total time spent in DL (standard deviation is 8.8 minutes).  On average, the students spend about 40 minutes of the 140 working with the whole class and 95 minutes working in small groups.  We find that the student experience in the 140-minutes DL sections varies vastly between different observations, with some classes spending as many as 91 minutes working as a whole class and some as few as 10 minutes (standard deviation of 20 minutes).  The remainder of the time is spent with students working in small groups, and varies from a minimum of 42 minutes to a maximum of 130 minutes.  In other words, some class periods are predominately spent working in a (whole class) group of 25-30 students with the instructor coordinating, whereas others are spent predominately in small groups (4-5 students) with hardly any time as a whole class. 
 
 \begin{figure*}[t!]
 \subcaptionbox{\label{fig:minSG}}
 {\includegraphics[height=.25\textheight]{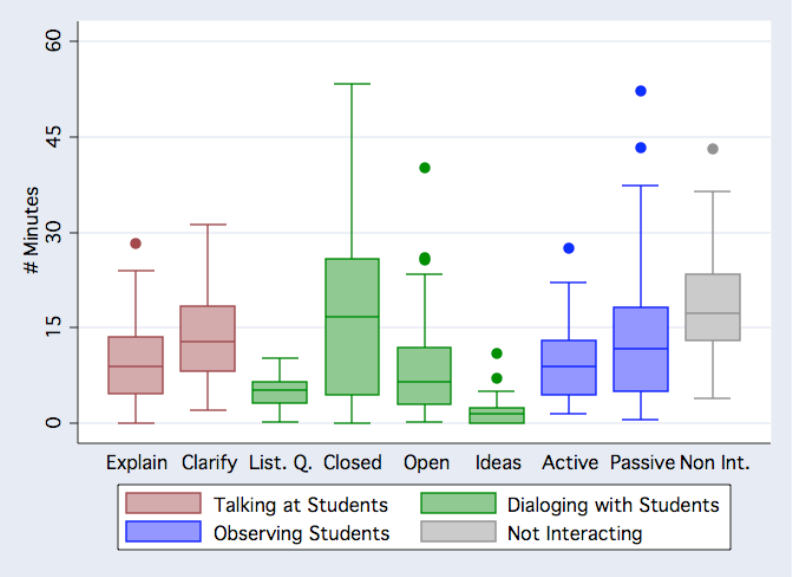}}
  \subcaptionbox{\label{fig:fractSG}}
  {\includegraphics[height=.25\textheight]{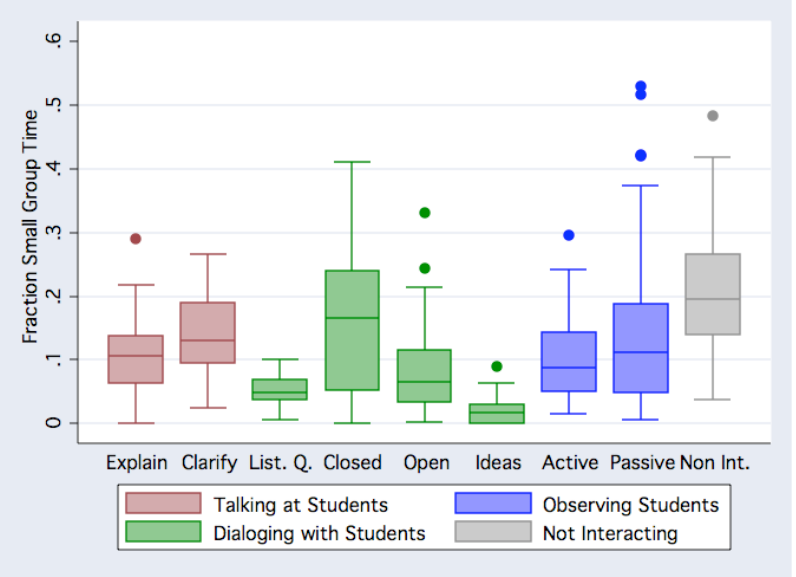}}
 \caption{Breakdown of small group time.  Left plot shows the number of minutes of small group time dedicated to each type of interaction, the right plot shows the percentage of small group time in each type of interaction.  Note that the category ÒideasÓ only contains 35 observations, because we added the category late.  In the other 24 observations all ÒideasÓ were categorized as Òopen-dialogue.Ó}
 \label{fig:SGtime}
 \end{figure*}

When we look at the WCD time allocation results by course, we immediately see large differences between CLASP A, B \& C.  Figure~\ref{fig:WCSGtime}b clearly illustrates that CLASP A students, on average, spend considerably more time in WCD mode than the CLASP B or C courses. Indeed, the CLASP A class with the least minutes spent in WCD mode (25 minutes) is similar to the CLASP C class with the most minutes spent in WCD mode (26 minutes), and the 25th percentile for WCD in CLASP A matches the 75th percentile for WCD in CLASP B.  The overall amount of time spent in DL is similar across all three courses in the series, though divided differently between small group and whole class time.

\subsection{Small Group Time}
We next turn our attention to the types of instructor-student interactions occurring during small group time.  As the overall number of minutes of small group time varies tremendously from observation to observation, results are presented both in terms of the number of minutes instructors spend in certain types of interactions and the percentage of overall small group time instructors spend in certain types of behaviors.  
 
We find that on average, instructor time is divided roughly evenly between talking at students (24\%), dialoguing with students (31\%), observing students (25\%) and not interacting with students (20\%) (see Figure~\ref{fig:SGtime}).  However, there is once again large variation across observations.  The subcategory of time spent explaining physics content to students ranges from as little as 0 minutes to as high as 28 minutes.  Likewise, some instructors spend hardly any time observing students working independently (as low as 4 minutes) whereas others spend as much as 54 minutes.  The data are presented both in terms of raw minutes spent engaging in an interaction type and percentage of overall small group time spent engaging in an interaction type.  The patterns are similar through either lens, though it should be remembered that the person with the most (or least) raw time spent participating in a certain interaction is not necessarily the person with the highest (or lowest) percentage of time.  The amount of time and proportion of time the instructor spends interacting with students in a way that the students share in the dialog also has a large range, from as low as 1.2 minutes to as high as 77 minutes, or in terms of percent, from 2.7\% to 63\%.  Not only does the amount of time students spend working in small groups vary, the type of instructor interaction the students experience also spans a large range.

 \begin{figure*}[t!]
  \subcaptionbox{\label{fig:minWC}}
 {\includegraphics[height=.25\textheight]{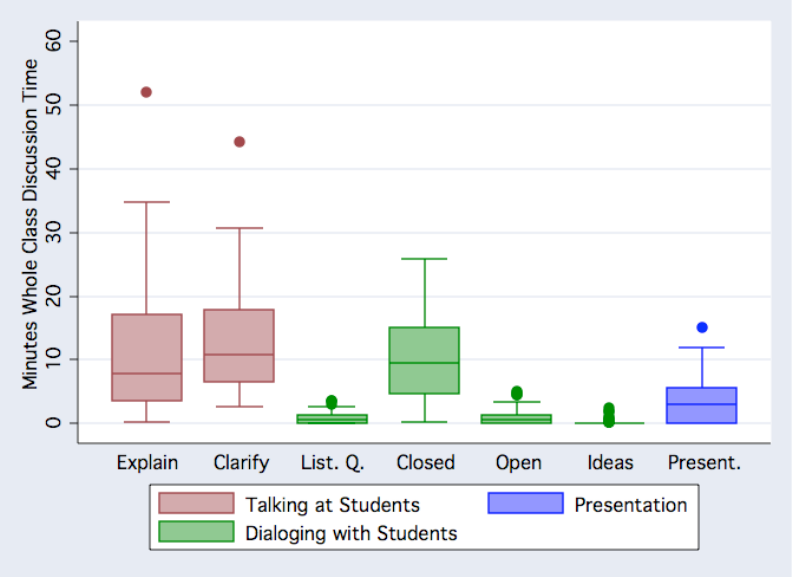}}
  \subcaptionbox{\label{fig:fractWC}}
  {\includegraphics[height=.25\textheight]{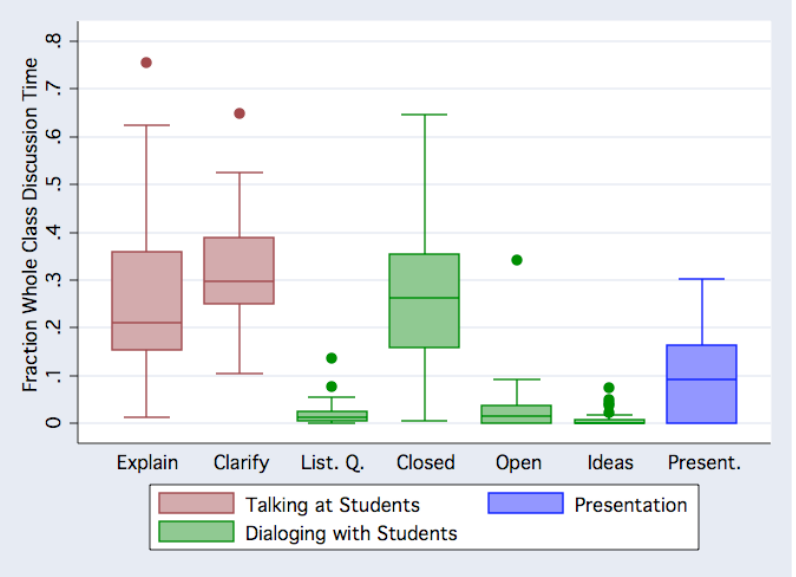}}
 \caption{Breakdown of whole class discussion time.  Left plot shows represents the number of minutes of WCD time dedicated to each type of interaction; the right plot shows the percentage of WCD time in each type of interaction.}
 \label{fig:WCtime}
 \end{figure*}
 
\subsection{Whole Class Discussion Time}
The goal of whole class discussion is to provide students with an opportunity to synthesize information and begin solidifying the work they did in their small groups.  Like during small group time, instructors use a variety of approaches to facilitate whole class discussion (WCD).  One technique is to have a student or group of students present the findings from the activity.  After the presentation, often the instructor summarizes/rephrases the student presentation, or asks other students to comment.  Another approach is for the instructor to reframe the questions in the activity sheet in such a way that only a single word or short series of words is required to give the answer, and pose these questions to the class in call and response format.  Sometimes the instructor will explain the ideas behind the questions to the students; other times the instructor will retain the short-answer format (which we refer to as closed dialogue) to get the students to fill in key aspects of the overarching ideas.  A third approach is for the instructor to convert the discussions into mini-lectures, with the instructor predominately explaining to the students.  A breakdown of the whole class discussion time by raw minutes and by percentage is shown in figure~\ref{fig:WCtime}.  

As with small group time, we see tremendous variation between instructors during WCD.  The most common interactions are explaining, clarifying, and dialoguing in closed form.  All instructors spend some time clarifying, which is unsurprising considering that this category was used to mark transitions between small group time and whole class time, in addition to coding actual clarification.  Other than clarification, each of the other modes were never used by some instructors and frequently used by others.

\subsubsection{}
\section{Discussion}
\subsection{RIOT as a research tool}
The RIOT was specifically developed using observations from CLASP courses.  For this reason, we are not sure the categories we identified will be as useful in other classroom environments. However, the RIOT has provided us with invaluable information about what is happening in our classrooms.  Researchers interested in using the RIOT for their own efforts might first ask themselves if the categories presented here effectively capture what is being looked for.  We recognize that the interactions may look different in other environments and that certain categories might need to be added or removed for different types of classrooms.  Our intention is not to showcase an instrument that could eventually be used to measure if certain classrooms meet a 'standard' of interactivity, but rather to introduce one that illustrates the pedagogical landscape of the classroom for the purposes of internal reflection.  This is not to say that the RIOT could not eventually be used as a metric for interactivity (in the way RTOP is similarly used), but since we have only observed a single curriculum using RIOT, it has not yet been verified that this is an effective use of the tool.

\subsection{Interactions from the perspective of the curriculum developer}
Although the CLASP course philosophy expects certain actions of the TAs, we find a vast range of instructor-student interactions within our interactive-engagement course.  The TAs are well-exposed to the course philosophy throughout their professional development and the curricular materials include prompts in the instructor notes meant to reinforce the philosophy, but nonetheless the actual implementation is extremely varied.  It is important to note that we did not expect the instructors to interact identically (nor would we want them to) and that, at this point, we are unable to identify which behaviors, if any, are ideal.  

For example, excessive 'explaining' is discouraged in TA training, because students are not necessarily engaged during this interaction, and yet there are some instructors who spend the vast majority of their whole class time doing just that. (See B in Figure~\ref{fig:7Bobs2}, Na in Figure~\ref{fig:7Aobs1}, \& H in Figure~\ref{fig:7Bobs2} in the Appendix).  Additionally, since the students are prompted in the instructor notes to have whole class discussions at specific points in the activities, we did not anticipate that some instructors would omit most of these discussions (see F, Cu \& Li in Figure~\ref{fig:7Bobs2} in appendix) while others would spend the majority of time in whole class discussion mode (see Mg, N , \&V in Figure~\ref{fig:7Aobs1}, in appendix).  Additional research should be undertaken to see if these deviations from expectation are helpful or harmful to the students.

We do not believe that the instructors purposely ignore their training to teach in an alternative method.  Instead the instructors incorporate the training they receive into their existing framework for teaching, which is typically the result of traditional physics instruction. Each individual TA interprets the curricular goals (as communicated through TA training, twice weekly TA meetings, and instructor notes) in a unique manner, and combines it with their prior experience leading to unique application in the classroom.  This observation is consistent with the constructivist teaching philosophy that guides the instruction of students in CLASP, so it is unsurprising to find the same type of learning in the instructor population.

At this point, we are not making judgments about what constitutes good or bad teaching interactions.  It is possible that the CLASP curriculum is sufficiently robust (i.e.  the students have enough opportunities to participate in an interactive environment with their peers) that the individual instructor interactions do not matter.  To make such a judgment we would need to either attempt to correlate these interactions with student achievement and we would need to have a better qualitative understanding of the immediate interaction effect on the student.  The latter is discussed in the remainder of this section.  

\subsection{Rethinking interactions from the student perspective}
Because an instructor can only spend limited time with each student or group without ignoring other students, a successful student-instructor interaction should be one where students are able to continue thinking about and making sense of phenomena after the instructor leaves the group.  This means that the method used to incite student thinking (which we have captured) is not as important as the resulting student reaction (which we have not captured).  It is possible that different instructors use different techniques to procure the same result, and if this result is Ôstudent sense-makingÕ any of those methods are viable.  Instead of only coding ÒexplainÓ or Òopen dialog,Ó it would be insightful to code the instructor-student interaction and the resulting student reaction to generate a log of which types of interactions lead to the types of thinking desired.  We did not use this approach because it would have required many more observers (there are more students to watch than instructors), and would not have allowed us to observe as many instructors, but this could be the subject of future research. 

We have little data on the student actions (student-student interactions) while the instructor is with other groups.  When the instructor is observing, we know that all groups in the class are working independently.  We should keep in mind that even while the instructor is interacting with a group, about 80\% of the class remains working without the instructor.  The actions the students take while the instructor is occupied with other groups are certainly important.  We suspect that the instructor-student interactions and the resulting classroom culture influence the actions of students while they are not with the instructor, though this data has not been captured.  It seems plausible that students who expect explanations when the instructor arrives at their group will behave differently than students who expect to have no interaction with the instructor and differently than students who expect to be prompted by a question from the instructor once the instructor arrives at their group.  The observers in this study noted that it seemed like instructors who explained to their students during small group time were more likely to have students wait for them to come help than to have students who continued trying to work through the more difficult parts of the problem.  Unfortunately, because of the nature of the study, no data other than anecdotal were taken in this regard. We have directly observed the instructor interactions, but have not captured data on overall student actions while not in the presence of the instructor; these actions also likely contribute to the classroom culture and student learning opportunities.

\section{Conclusions}
The instructor-student interactions in our physics course vary tremendously. The extent of the variations is surprising given that the graduate student TAs receive two intensive days of training, 10 weeks of continuing TA professional development, and extensive TA notes supporting a particular pedagogy.

The vast interaction differences should be considered when planning curricula that depend on multiple individuals, such as graduate students, for implementation. The instructors responsible for enacting the ÒreformedÓ aspects of the curriculum are not always the same individuals who created the course, so it is important to know to what extent their enactment matches the goals of the course creators.  Similarly, it is important for instructors who have enacted curricula developed elsewhere to understand the goals of the curricula used; there is a high likelihood that some percentage of the enactment differs from the vision of the creators.

Instructors are different, and we can depend on them to be so.  Graduate student instructors turn over rapidly, teaching for no more than a couple of years. Therefore, even in a longstanding reformed course, with an extensive TA training course, large variations remain in instructor implementation.  At this point, we are describing the existence and extent of the variation and not trying to determine why some categories contain more variation than others.  The range of instructor interactions ensures that students in different sections of the course experience the course differently.  This may not be a problem depending on the resulting student reaction to the interaction.  However, we hypothesize that it is possible that explaining (as we have categorized it) fails to promote active sense-making from students in the same way that traditional lecture does.  Therefore, in order to ensure students in each class have similar learning opportunities, we either need to write these opportunities more explicitly into the curriculum, or TAs need more explicit and personalized TA training in their first quarter of teaching.

\subsection{}
\subsubsection{}

\begin{acknowledgments}
This research was funded in part by the Teaching Resources Center (now the Center for Excellence in Teaching and Learning) at UC Davis.  
\end{acknowledgments}

\bibliographystyle{apsrev4-1.bst}
\bibliography{InteractionsWest2.bib}

\appendix
\section{Screen shot of RIOT interface}
\begin{figure}[H]
\includegraphics[height=.65\textheight]{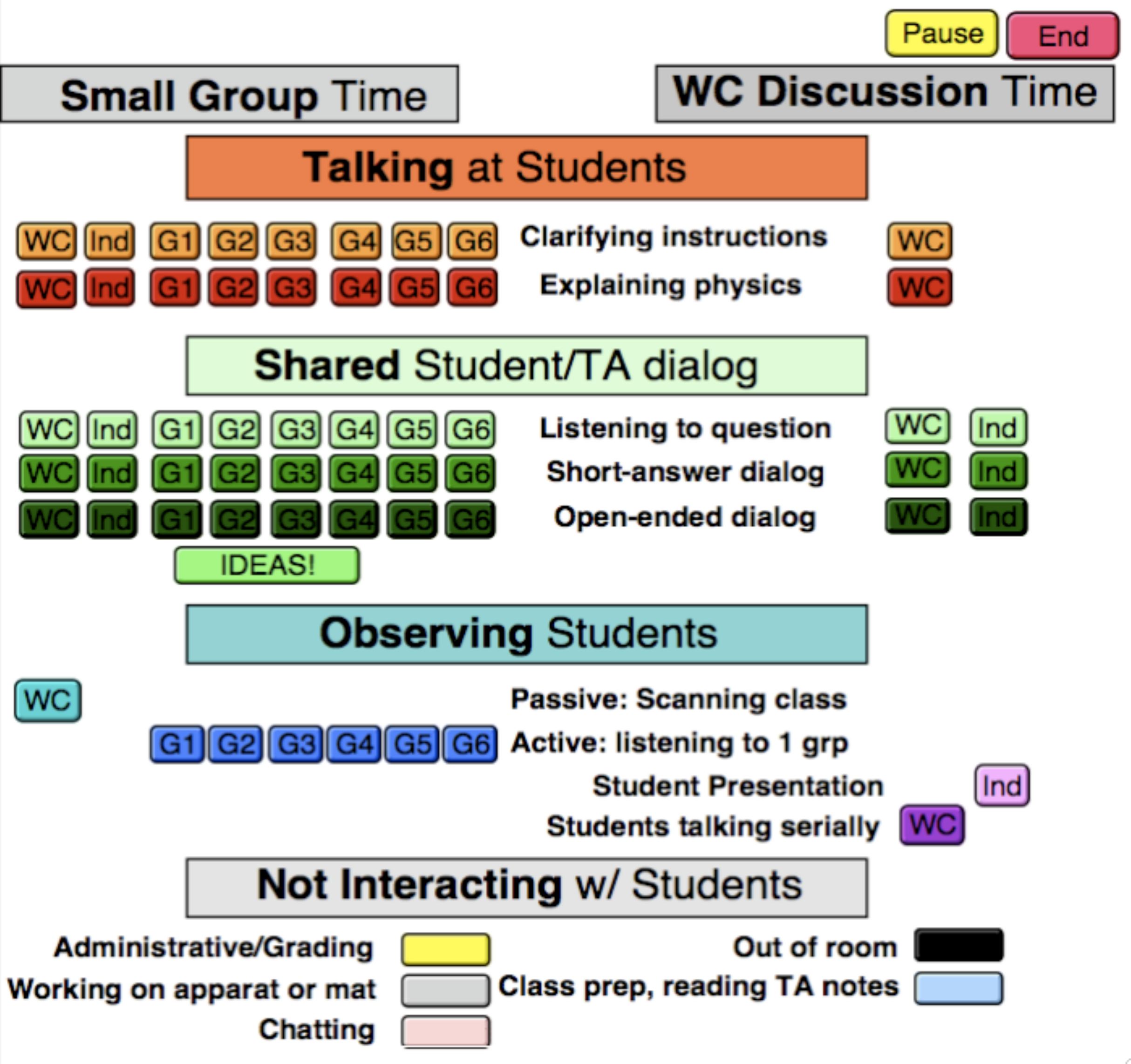}
\caption{Screen shot of the Real-time Instructor Observing Tool (RIOT): The left side of the screen is for when the class is in Small Group mode, and the right side is for when the class is in Whole Class mode. The screen is divided horizontally into four larger categories:  Talking at Students, Talking with Students, Observing Students, and Not Interacting with Students.  Under each of these larger categories are more precise categories.  Each row of icons corresponds to the smaller category name that is written next to it.  Each small icon labeled WC, Ind, and G1-G6, represent who the instructor is interacting with (Whole Class, an Individual, and Groups1-6 respectively).}
\label{fig:screenshot}
\end{figure}

\section{Twelve Different Instructors Teaching the same set of Activities }
\begin{figure}[H]
\includegraphics[height=.5\textheight]{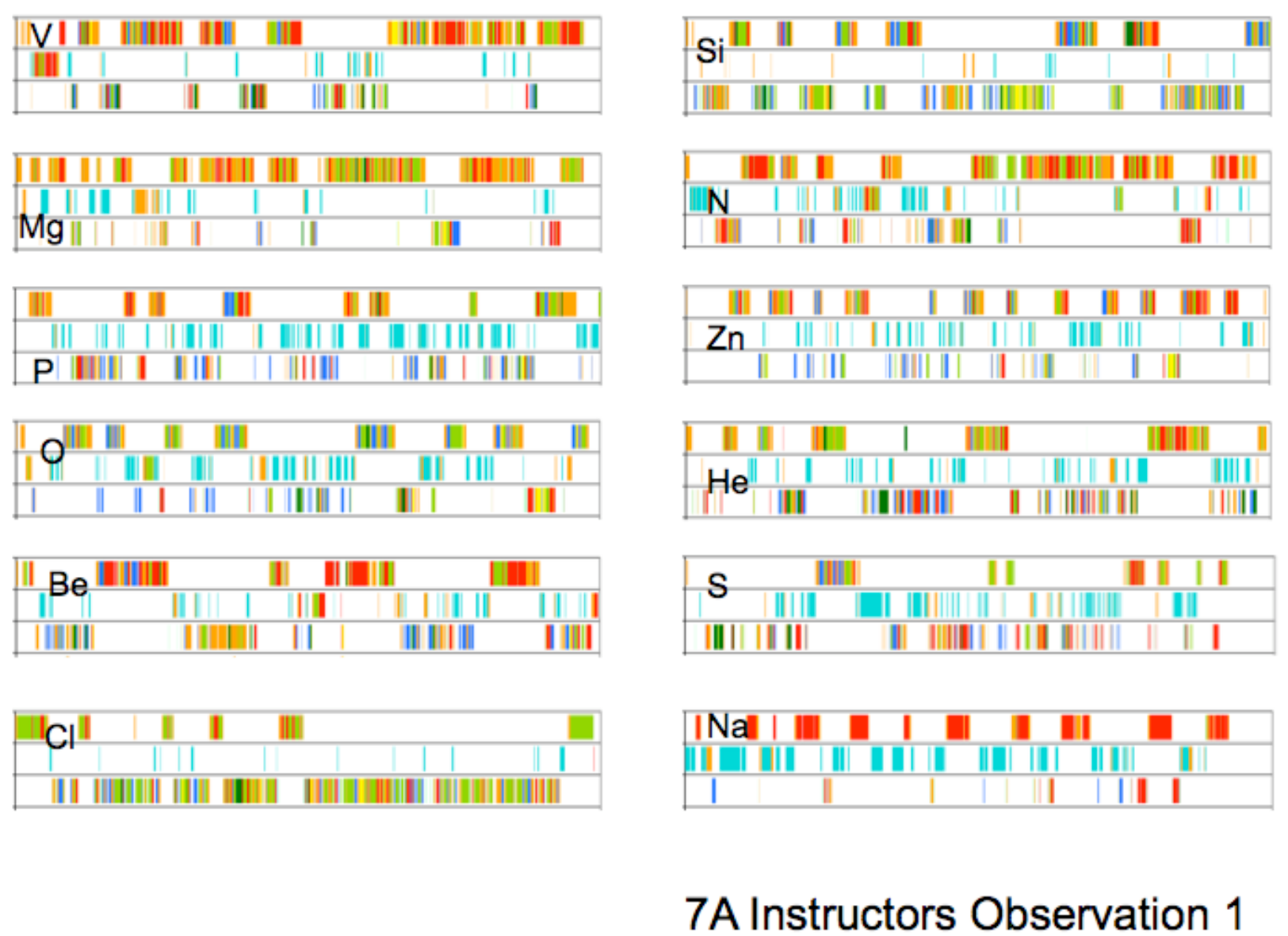}
\caption{The top row represents Whole Class time, the bottom row is Small Group time, and the middle row represents the interations that take place during Small Group time, but when the instructor is addressing the class as a whole. Time is on the y-axis.  The colors are consistent with those in figure~\ref{fig:T3}.}
\label{fig:7Aobs1}
\end{figure}

\section{Eleven Different Instructors Teaching the same set of Activities}
\begin{figure}[H]
\includegraphics[height=.5\textheight]{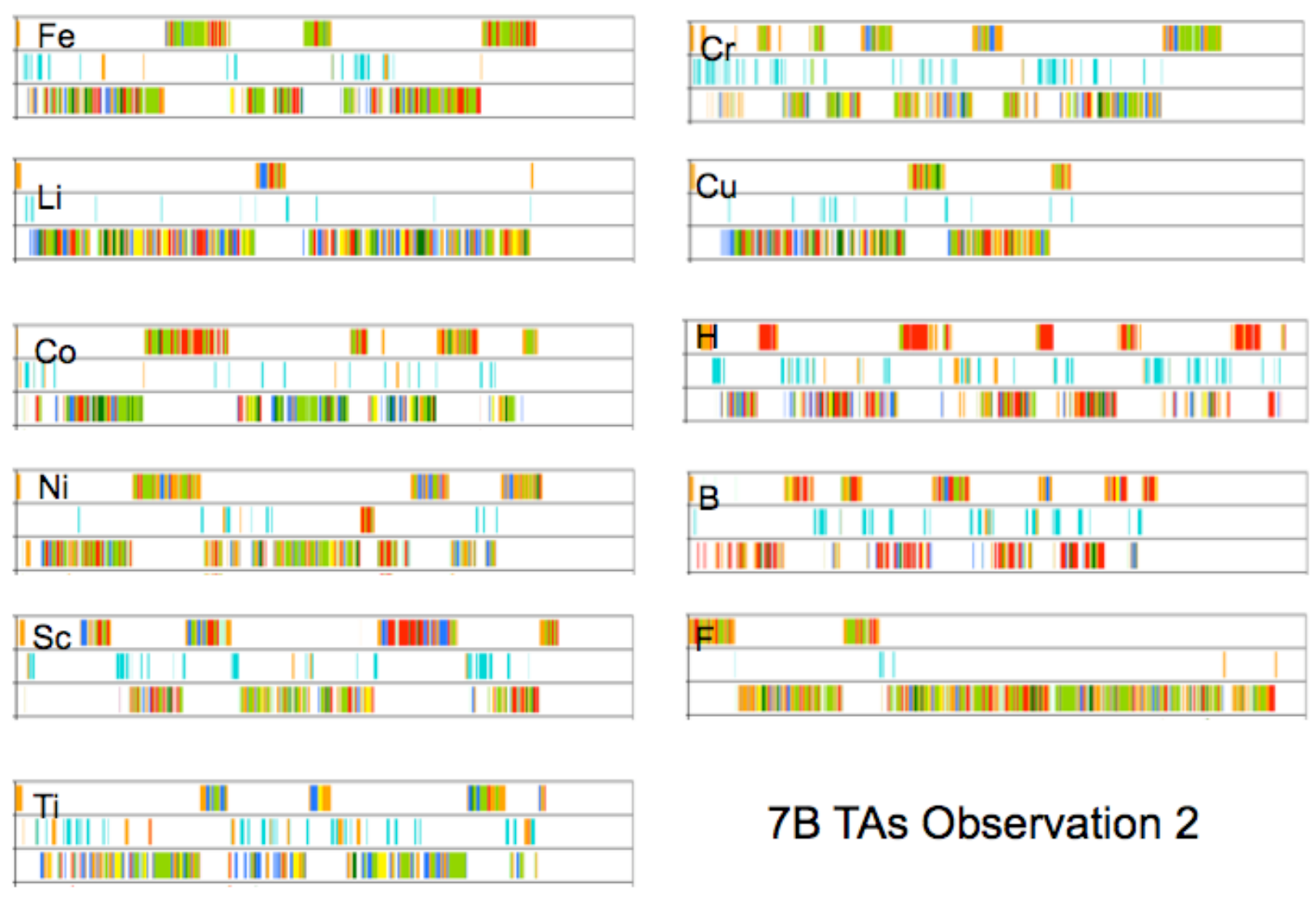}
\caption{The top row represents Whole Class time, the bottom row is Small Group time, and the middle row represents the interations that take place during Small Group time, but when the instructor is addressing the class as a whole. Time is on the y-axis. The colors are consistent with those in figure~\ref{fig:T3}.}
\label{fig:7Bobs2}
\end{figure}

\end{document}